\documentclass[aps,prl,preprint,floatfix,superscriptaddress]{revtex4-1}
\usepackage{graphicx,txfonts,bm}
\usepackage[colorlinks,citecolor=magenta,linkcolor=blue]{hyperref}

\begin{document}

\title{Collapse of quasiparticle multiplets and $5f$ itinerant-localized crossovers in cubic phase Pu$_{3}$Ga}

\author{Li Huang}
\email{lihuang.dmft@gmail.com}
\affiliation{Science and Technology on Surface Physics and Chemistry Laboratory, P.O. Box 9-35, Jiangyou 621908, China}

\author{Haiyan Lu}
\affiliation{Science and Technology on Surface Physics and Chemistry Laboratory, P.O. Box 9-35, Jiangyou 621908, China}

\date{\today}

\begin{abstract}
The physical properties of plutonium and plutonium-based intermetallic compounds are extremely sensitive to temperature, pressure, and chemical alloying. A celebrated example is the high-temperature $\delta$ phase plutonium, which can be stabilized at room temperature by doping it with a few percent trivalent metal impurities, such as gallium or aluminum. The cubic phase Pu$_{3}$Ga, one of the plutonium-gallium intermetallic compounds, plays a key role in understanding the phase stability and phase transformation of the plutonium-gallium system. Its electronic structure might be essential to figure out the underlying mechanism that stabilizes the $\delta$ phase plutonium-gallium alloy. In the present work, we studied the temperature-dependent correlated electronic states of cubic phase Pu$_{3}$Ga by means of a combination of the density functional theory and the embedded dynamical mean-field theory. We identified orbital selective 5$f$ itinerant-localized (coherent-incoherent) crossovers which could occur upon temperature. Actually, there exist two well-separated electronic coherent temperatures. The higher one is for the $5f_{5/2}$ state [$T_{\text{coh}}(5f_{5/2}) \approx 700$~K], while the lower one is for the $5f_{7/2}$ state [$T_{\text{coh}}(5f_{7/2}) \approx 100$~K]. In addition, the quasiparticle multiples which originate from the many-body transitions among the $5f^{4}$, $5f^{5}$, and $5f^{6}$ electronic configurations, decay gradually. The hybridizations between the localized 5$f$ bands and conduction bands are subdued by high temperature. Consequently, the Fermi surface topology is changed, which signals a temperature-driven electronic Lifshitz transition. Finally, the calculated linear specific heat coefficient $\gamma$ is approximately 112~mJ / (mol~K$^2$) at $T = 80$~K, which suggests that the cubic phase Pu$_{3}$Ga might be a promising candidate of plutonium-based heavy-fermion system.   
\end{abstract}

\maketitle

\section{introduction}

In the actinides, the $5f$ shell is successively filled. With increasing atomic number, the $5f$ electrons exhibit a tantalizing and Janus-faced behavior (being itinerant or localized), which is no doubt responsible for a plethora of interesting lattice properties of the actinides~\cite{RevModPhys.81.235}. For the early actinides (from Th to Np), the $5f$ electrons incline to be itinerant and contribute to chemical bonding. They usually produce very narrow and flat energy bands, leading to particularly high density of states near the Fermi level~\cite{nature:1995}. Since lattice distortions can split these narrow $5f$ energy bands and thereby lower the total energy of the system effectively, it is not surprised that the ground states of the early actinides favor the low-symmetry crystal structures. On the contrary, the $5f$ electrons in the late actinides (from Am to No) tend to be localized and become chemically inert, giving rise to nontrivial local magnetic moments and nearly fixed atomic volumes~\cite{PhysRevLett.85.2961}. Plutonium, the sixth member of the actinide series (atomic member: 94, chemical symbol: Pu), separates the early and late actinides. Hence, its $5f$ electrons just live at the brink between localized and itinerant configurations, which may be unique in the periodic table. Since electronic structure determines nonnuclear properties, plutonium and its compounds naturally exhibit a barrage of unusual properties that we still don't fully understand so far~\cite{albers:2001}. Next we would like to illustrate some of them.

Pu is considered to be an element at odds with itself. Some people even claimed that it is one of the most mysterious and complicated elements in the periodic table~\cite{LAReview}. From the viewpoint of crystallography, there are six stable allotropes ($\alpha$, $\beta$, $\gamma$, $\delta$, $\delta'$, and $\epsilon$ phases) of plutonium at ambient pressure. Thus, pure Pu metal may undergo five successive solid state phase transitions before reaching its liquid phase~\cite{Hecker2004}. The crystal structures of the low-temperature $\alpha$ and $\beta$ phases are monoclinic, with eight and seven distinct lattice sites, respectively~\cite{Zachariasen:a03908,Zachariasen:a02472}. Note that such low-symmetry structures are unique to Pu among elements~\cite{zhu:2013,PhysRevB.99.125113}. For the high-temperature $\delta$ and $\epsilon$ phases of Pu, the crystal symmetries become higher. For instance, the $\delta$-Pu has a face-centered cubic structure, but it exhibits the largest atomic volume among the six stable phases of Pu. As a result, the $\alpha-\delta$ structural transition is accompanied with a huge volume expansion ($\approx 25$\%). Applying small pressure to the $\delta$ phase, it is destabilized and replaced with a new $\zeta$ phase, whose crystal structure remains secret heretofore. While the $\alpha$ phase would transform into a hexagonal phase when the pressure is around 40 GPa~\cite{Hecker2004}. From the viewpoint of lattice properties, plutonium, especially its $\delta$ phase, is notoriously peculiar~\cite{LAReview}. The brittle $\alpha$ phase could expand at a rate almost five times the rate in iron when heated. However, the ductile $\delta$ phase would contract while being heated, which means that the $\delta$ phase has a negative thermal expansion coefficient. The magnetic behavior of Pu is a long-standing puzzle in the condensed matter physics community. Its low-temperature magnetic susceptibility is extraordinary high and almost keeps a constant value, which signals some sort of magnetism. However, for any phase of Pu, no long-range magnetic ordering has been observed even at the lowest temperature~\cite{Janoscheke:2015,PhysRevB.72.054416}. The electrical resistivity of the $\delta$ phase is an order of magnitude larger than those of the other simple metals at room temperature. It even increases steadily as the temperature is lowered to 100~K. Furthermore, the Sommerfeld coefficient of the specific heat of $\delta$-Pu is larger than that of any element. It is large enough so that some people argued that it should be classified as a heavy-fermion system~\cite{PhysRevLett.91.205901}. The phonon dynamics and elastic properties of $\delta$-Pu are also quite exotic~\cite{dai:2003,wong:2003}. The phonon dispersion curves of $\delta$-Pu shows a pronounced softening along the transverse acoustic branch (i.e. the $T[111]$ mode), which implies its lattice instability and the Bain path for the $\delta-\epsilon$ structural transition~\cite{PhysRevLett.92.146401}. The nearly degenerated longitudinal and transverse acoustic phonon branches along the $[001]$ direction lead to approximately equal elastic constants $C_{11}$ and $C_{44}$, and astonishing shear anisotropy ($C'$). Overall, the six stable allotropes of Pu are virtually different metals. Their $5f$ electronic structures should be completely diverse~\cite{PhysRevB.101.125123,Brito:2020,PhysRevX.5.011008}.

Since the $5f$ electron wave-functions show large spatial extent and have large overlap with the $spd$ conduction electron wave-functions (i.e., the $5f$ electrons are able to participate in chemical bonding)~\cite{LAReview}, Pu can constitute a variety of intermetallic compounds with the other elements in the periodic table~\cite{RevModPhys.81.235}. Just like the pure Pu, the Pu-based intermetallic compounds also demonstrate rich physical phenomena, including complex magnetic ordering states, valence state fluctuation, heavy-fermion behaviors, quantum criticality, and unconventional superconductivity, just to name a few~\cite{bauer_review,PhysRevLett.108.017001,Sarrao2002}. They also span the range from itinerant to localized $5f$ electronic behavior. The first example is about the Pu-based ``115'' heavy-fermion superconductors, Pu$M$$X_{5}$ (where $M =$ Co, Rh; $X =$ Ga, In). These compounds are isostructural, crystallizing in the tetragonal HoCoGa$_{5}$-type structure. But, the volume collapse between PuCoIn$_{5}$ and PuCoGa$_{5}$ reaches 25\% (similar to the volume decrease between $\delta$-Pu and $\alpha$-Pu), indicating that the $5f$ electrons in PuCoIn$_{5}$ are more localized than those in PuCoGa$_{5}$~\cite{zhu:57001}. These materials have attracted quite a lot of research interests, because they evince non-Fermi-liquid behavior and close to antiferromagnetic quantum critical point~\cite{PhysRevLett.108.017001}. Furthermore, the superconducting critical temperature $T_c$ of PuCoGa$_{5}$ is about 18.5~K, which is the highest for the heavy-fermion $f$-electron systems~\cite{Sarrao2002,Curro2005,Daghero2012}. Another example concerns with the PuB$_{x}$ (where $x =$ 2, 4, 6, and 12) system. PuB$_{2}$, PuB$_{4}$, PuB$_{6}$, and PuB$_{12}$ crystallize in the hexagonal AlB$_{2}$-type, tetragonal UB$_{4}$-type, cubic CaB$_{6}$-type, and face-centered-cubic structures, respectively. They are typical mixed-valent compounds. Both PuB$_{4}$ and PuB$_{6}$ are semiconductors. Recently, they are predicted to be promising candidates of strongly correlated topological insulators, in which the bulk states are insulating but the topologically protected surface states are metallic~\cite{PhysRevB.99.035104,PhysRevLett.111.176404}. 

Now let us turn to the Pu-Ga systems. Perhaps they are the most watched Pu-based materials due to their broad applications in the military and energy industries~\cite{LAReview}. As is mentioned before, $\delta$-Pu usually stabilizes under high temperature, but doping it with a few percent trivalent metal impurities (the most often used element is Ga) may stabilize it down to room temperature. According to the Pu-Ga phase diagram, the $\delta$ phase Pu-Ga alloy could be metastable over a wide range of temperature~\cite{Hecker2004}. Though the $\delta$ phase Pu-Ga alloy has been widely used, however, the underlying mechanism about why a few percent Ga impurities can delay the $\delta-\alpha$ phase transition and stabilize the $\delta$-Pu's lattice remains unknown and is warmly debated up to now~\cite{HECKER2004429,PhysRevB.72.205122,PhysRevB.81.224110,ROBERT2007191,PhysRevB.94.214108,PhysRevLett.96.206402,PhysRevLett.92.185702}. Besides the Pu-Ga alloy (disorder), the Pu-Ga system contains quite a few intermetallic compounds, including PuGa (tetragonal), PuGa$_{2}$ (hexagonal), PuGa$_{3}$ (rhombohedral), PuGa$_{3}$ (hexagonal), PuGa$_{4}$ (orthorhombic), PuGa$_{6}$ (tetragonal), Pu$_{2}$Ga$_{3}$ (hexagonal), Pu$_{5}$Ga$_{3}$ (tetragonal), Pu$_{3}$Ga (tetragonal), and Pu$_{3}$Ga (cubic), etc~\cite{HECKER2004429}. To our knowledge, most of these compounds have not been carefully studied with either theoretical calculations or experiments. We also notice that Pu$_{3}$Ga actually has two forms: a AuCu$_{3}$-type cubic phase [see Fig.~\ref{fig:akw}(a)-(b)] at room temperature and a SrSb$_{3}$-type hexagonal phase at low temperatures. The hexagonal phase is essentially a slight tetragonal deformation of the cubic phase (it is stretched along the $c$-axis). When the atomic percent of Ga in the $\delta$ phase Pu-Ga alloy is small ($< 5$\%), though the equilibrium process can be very slow, the decomposition into a mixture of $\alpha$-Pu and the cubic phase Pu$_{3}$Ga at low temperature is allowed thermodynamically~\cite{HECKER2004429,Hecker2004}. Quite recently, Zhang \emph{et al.} employed the density functional theory plus $U$ (static Coulomb interaction) method to study the electronic and vibrational properties of cubic phase Pu$_{3}$Ga~\cite{PhysRevB.96.134102}. They found that there exists strong hybridizations between the Pu-$5f$, Pu-$6d$ states and the Ga-$3d$, Ga-$4p$ states. Thus, the presence of Ga impurities strongly acts on the electronic structures of surrounding Pu atoms. More interestingly, they also observed significant phonon softening in the transverse acoustic branch at the $L$ point in the Brillouin zone. Therefore, they argued that the Ga atoms should be responsible for this phonon anomaly. Since similar behavior has been observed in the $\delta$ phase Pu-Ga alloy~\cite{wong:2003,PhysRevLett.92.146401}, they concluded that the change in electronic structures will eventually lead to the changes in lattice vibrations and phase stability of the $\delta$-Pu~\cite{PhysRevB.96.134102,PhysRevB.100.184101}. This viewpoint sounds reasonable, but it is challenged by the fact that the phonon softening phenomenon in pure $\delta$-Pu has been predicted by theoretical calculations~\cite{dai:2003}, prior to any experiments~\cite{wong:2003,PhysRevLett.92.146401}. Anyway, it is obvious that the $5f$ electronic structure is a key to understand the structural, magnetic, transport, and lattice dynamics properties of plutonium and its intermetallic compounds. So, in order to shed new light into the phase transitions and phase stability of the $\delta$ phase Pu-Ga alloy, it is highly desired to capture and understand the $5f$ electronic structure of cubic phase Pu$_{3}$Ga at first~\cite{HECKER2004429,Hecker2004,qua.26105}. 

In the present work, we tried to study the evolution of correlated electronic structures of cubic phase Pu$_{3}$Ga upon temperature by means of a state-of-the-art first-principles many-body approach, namely the density functional theory in combination with the embedded dynamical mean-field theory (dubbed DFT + DMFT)~\cite{RevModPhys.78.865,RevModPhys.68.13}. We found that at low temperature the $5f$ electrons in the cubic phase Pu$_{3}$Ga are itinerant, with remarkable quasiparticle multiplets in the vicinity of the Fermi level. When the temperature is increased, orbital selective 5$f$ itinerant-localized crossovers will take place, accompanying with ``melt'' of the quasiparticle multiplets. During this electronic structure transition, the Fermi surface topology, the 5$f$ occupancy, and the hybridization gaps are modified accordingly.    

\section{Results}

\begin{figure*}[ht]
\includegraphics[width=\textwidth]{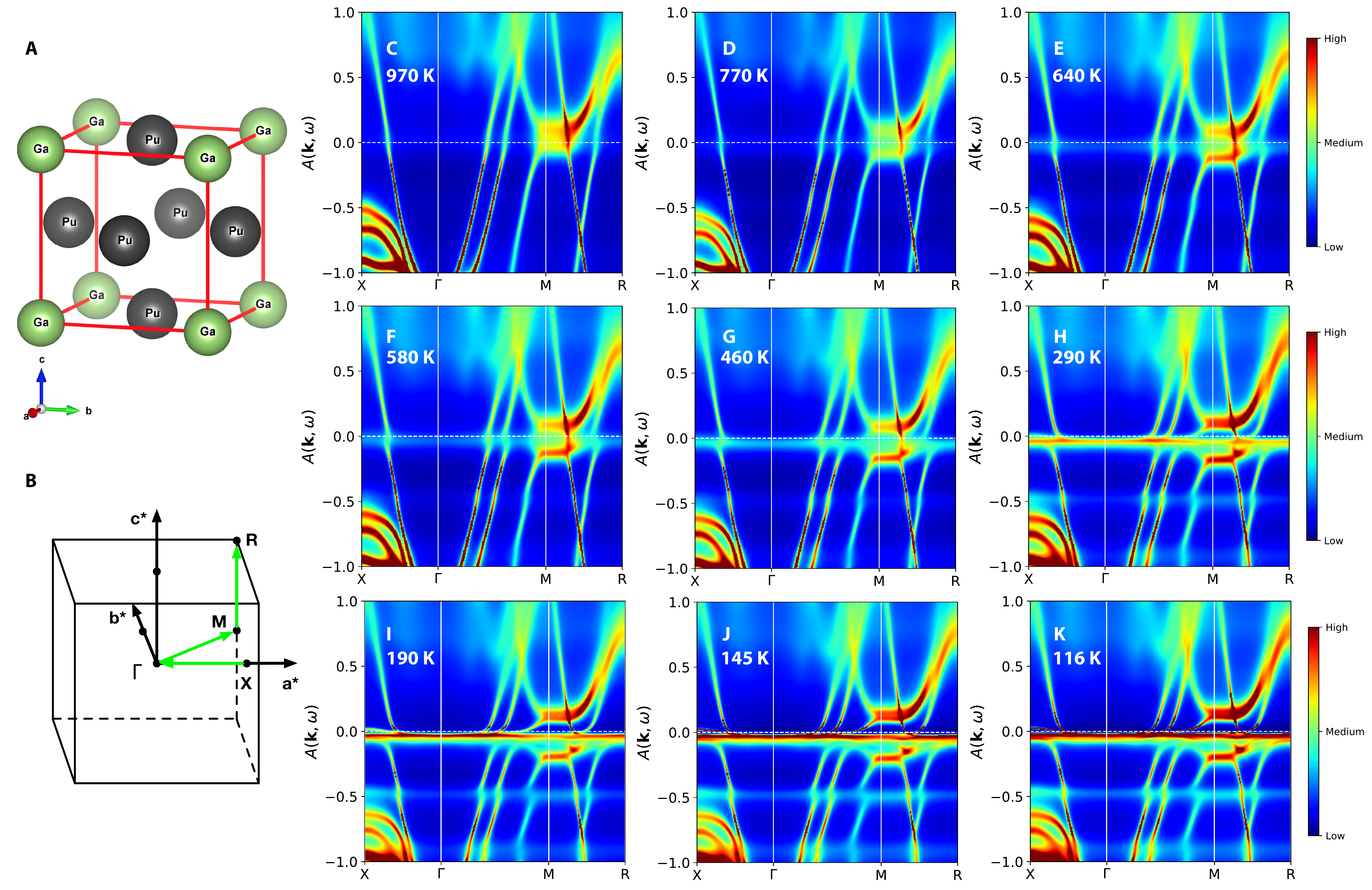}
\caption{\textbf{Crystal structure and temperature-dependent quasiparticle band structures of cubic phase Pu$_{3}$Ga.} \textbf{a} Schematic crystal structure of cubic phase Pu$_{3}$Ga. The Pu and Ga atoms are presented as grey and green balls, respectively. \textbf{b} The first Brillouin zone of cubic phase Pu$_{3}$Ga. The green arrows denote the high-symmetry directions ($X-\Gamma-M-R$) that used in the following band structure calculations. \textbf{c-k} DFT + DMFT quasiparticle band structures of cubic phase Pu$_{3}$Ga calculated at various temperatures. In these panels, the horizontal dashed lines denote the Fermi level. When $T \leq 640$~K, distinct quasiparticle bands and hybridization gaps appear in the vicinity of the Fermi level. \label{fig:akw}}
\end{figure*}

\begin{figure*}[ht]
\includegraphics[width=\textwidth]{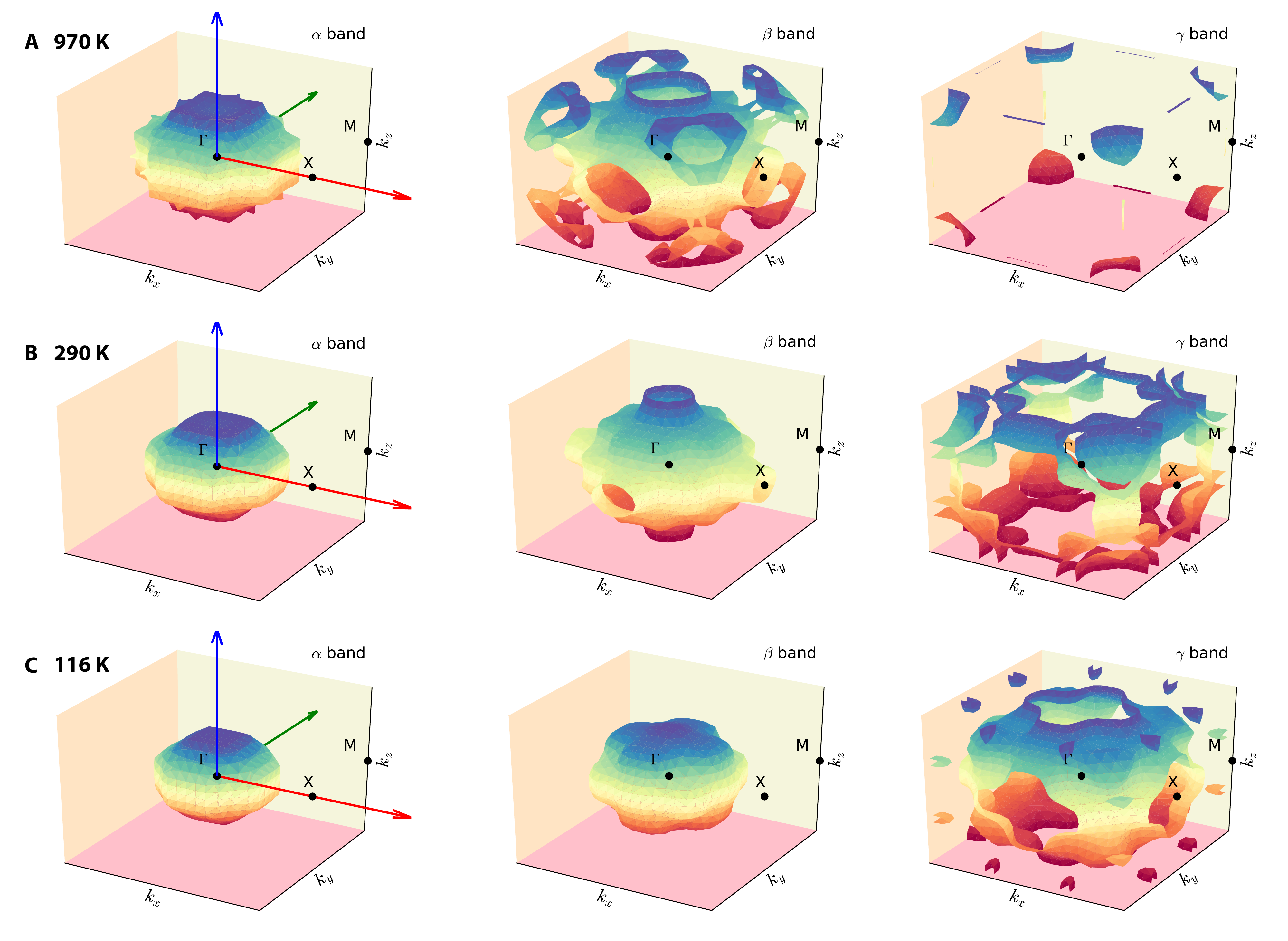}
\caption{\textbf{3D Fermi surfaces of cubic phase Pu$_{3}$Ga obtained by the DFT + DMFT method.} \textbf{a} $T = 970$~K. \textbf{b} $T = 290$~K. \textbf{c} $T = 116$~K. There are three doubly degenerated bands (labelled by $\alpha$, $\beta$, and $\gamma$) crossing the Fermi level. They are plotted in the left, middle, and right columns, respectively. \label{fig:fs3d}}
\end{figure*}

\begin{figure*}[ht]
\includegraphics[width=\textwidth]{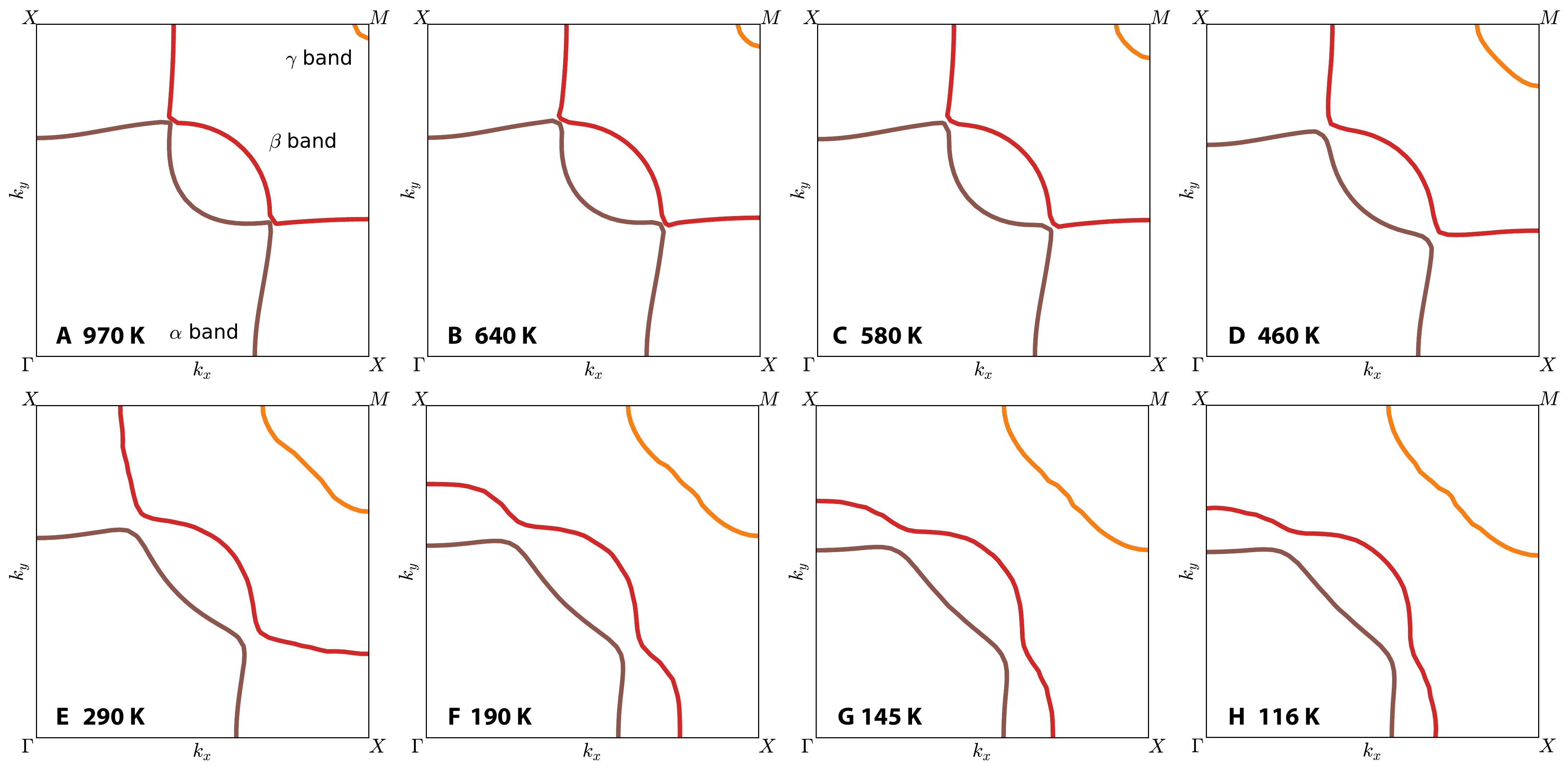}
\caption{\textbf{2D Fermi surfaces of cubic phase Pu$_{3}$Ga obtained by the DFT + DMFT method.} In this figure, only the $k_x-k_y$  plane (with $k_z = \pi/2$) is shown. There are three doubly degenerated bands (labelled by $\alpha$, $\beta$, and $\gamma$) that crossing the Fermi level. They are visualized using different colors. \label{fig:fs2d}}
\end{figure*}

\begin{figure*}[ht]
\includegraphics[width=\textwidth]{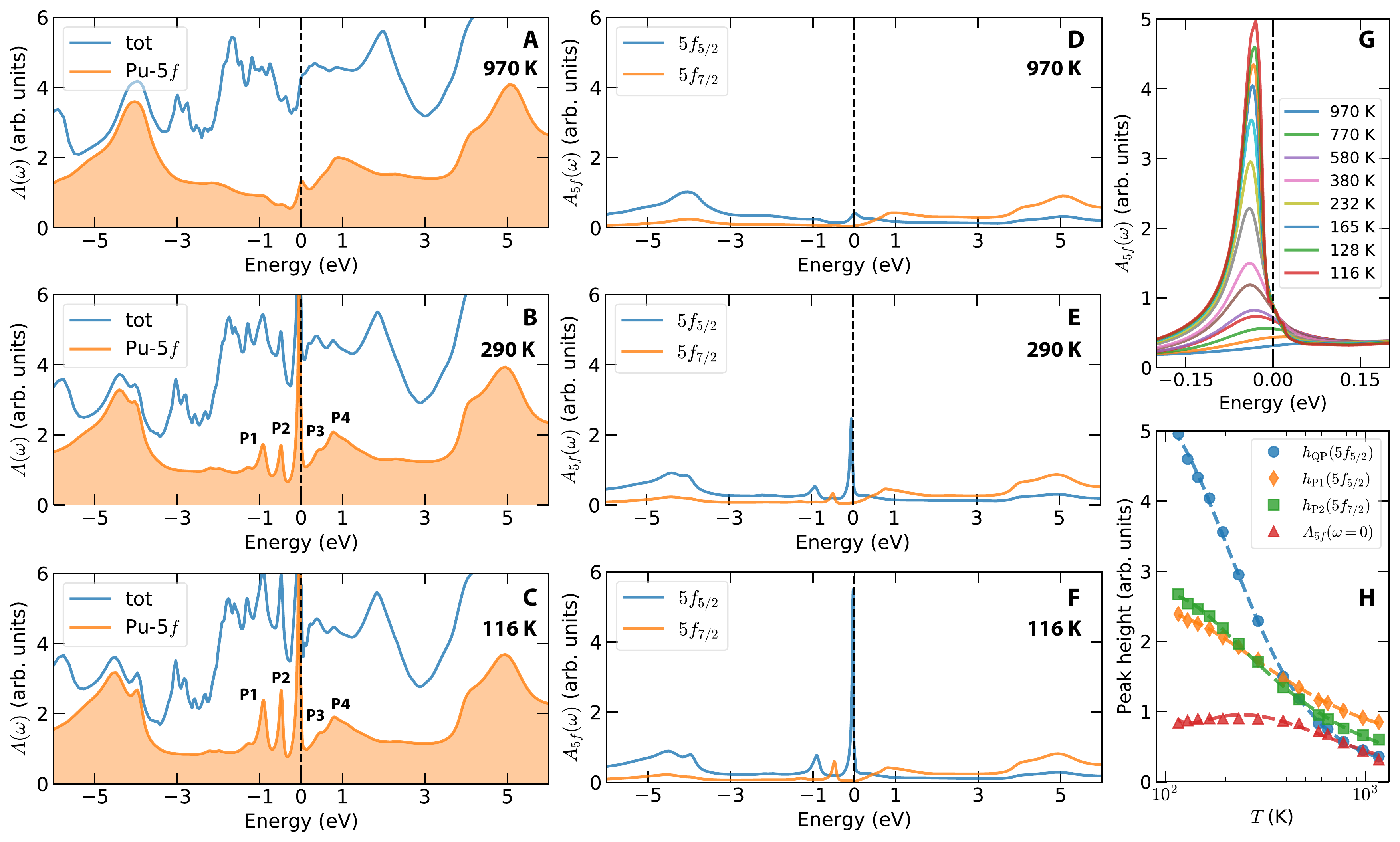}
\caption{\textbf{Density of states of cubic phase Pu$_{3}$Ga obtained by the DFT + DMFT method.} \textbf{a-c} Total density of states (solid lines) and $5f$ partial density of states (color-filled regions) of cubic phase Pu$_{3}$Ga at $T = 970$~K, $290$~K, and $116$~K. The peaks which stemming from the quasiparticle multiplets are annotated with ``P1'', ``P2'', ``P3'', and ``P4''. \textbf{d-f} The $j$-resolved $5f$ partial density of states [i.e. $A_{5f_{5/2}}(\omega)$ and $A_{5f_{7/2}}(\omega)$] of cubic phase Pu$_{3}$Ga at $T = 970$~K, $290$~K, and $116$~K. \textbf{g} Temperature-dependent central quasiparticle peaks (they are mainly comprised of the $5f_{5/2}$ states). From top to bottom, the system temperature $T$ increases gradually ($T$ is from 116~K to 970~K). \textbf{h} The height of the central quasiparticle peak $h_{\text{QP}}(5f_{5/2})$, the height of the quasiparticle multiplet peak ``P1'' $h_{\text{P1}}(5f_{5/2})$, the height of the quasiparticle multiplet peak ``P2'' $h_{\text{P2}}(5f_{7/2})$, and the $5f$ spectral weight at the Fermi level $A_{5f}(\omega =0)$ as a function of temperature $T$. \label{fig:dos}}
\end{figure*}

\begin{figure*}[ht]
\includegraphics[width=\textwidth]{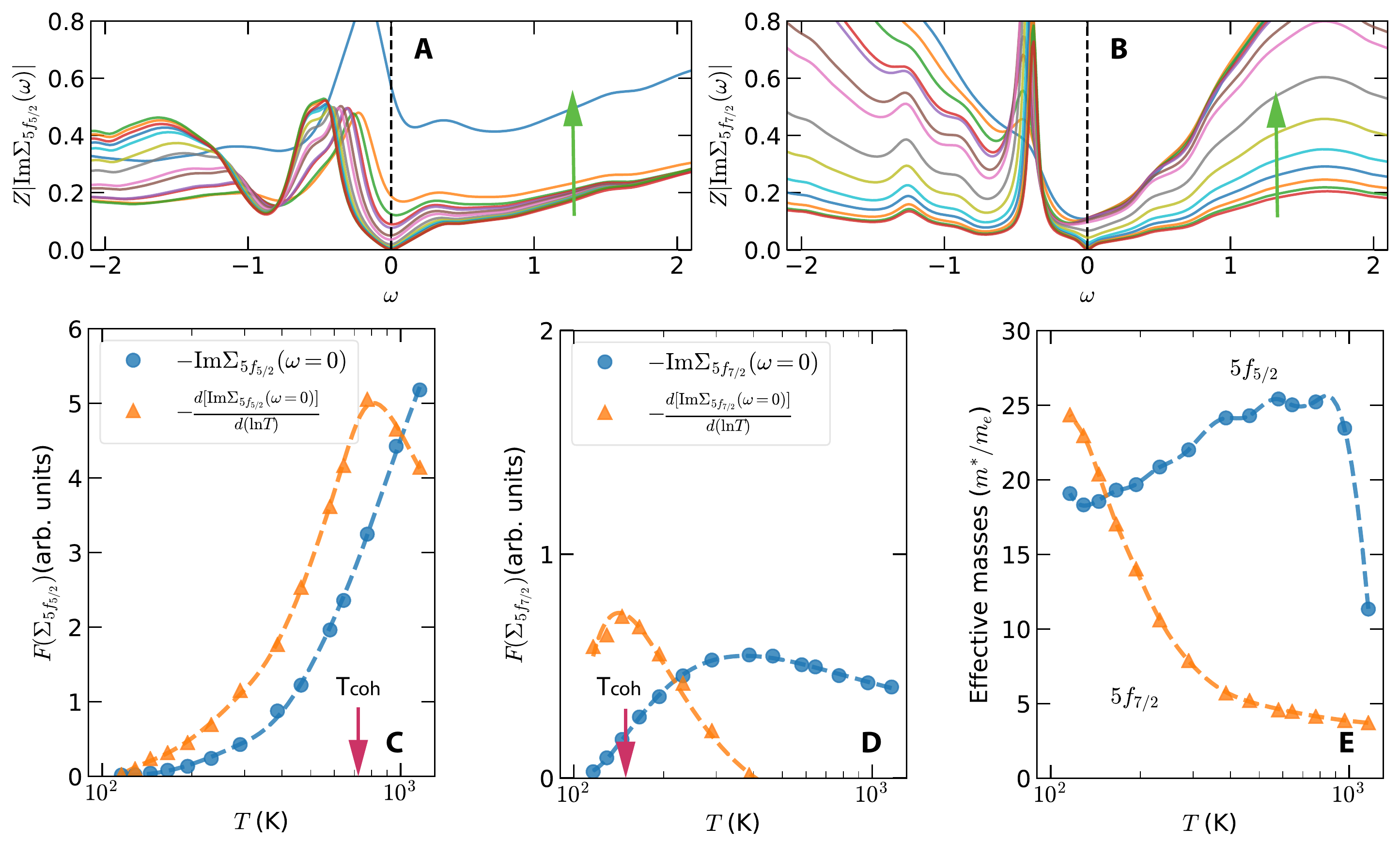}
\caption{\textbf{Real-frequency self-energy functions of cubic phase Pu$_{3}$Ga obtained by the DFT + DMFT method.} \textbf{a-b} Temperature-dependent $Z|\text{Im}\Sigma(\omega)|$ for the $5f_{5/2}$ and $5f_{7/2}$ states. $Z$ means the renormalization factor. In these panels, the green arrows denote the increasing system temperature ($T$ is from 116 K to 970 K). \textbf{c} $-d\text{Im}\Sigma_{5f_{5/2}}(\omega = 0) / d\text{ln}T$ and $-\text{Im}\Sigma_{5f_{5/2}}(\omega=0)$ as a function of temperature $T$. \textbf{d} $-d\text{Im}\Sigma_{5f_{7/2}}(\omega = 0) / d\text{ln}T$ and $-\text{Im}\Sigma_{5f_{7/2}}(\omega=0)$ as a function of temperature $T$. In panels (c) and (d), the red arrows indicate the coherent temperatures [$T_{\text{coh}}(5f_{5/2}) \approx 700$~K and $T_{\text{coh}}(5f_{7/2}) \approx 100$~K]. \textbf{e} Effective electron masses for the $5f_{5/2}$ and $5f_{7/2}$ states as a function of temperature $T$. \label{fig:sig}}
\end{figure*}

\begin{figure*}[ht]
\includegraphics[width=\textwidth]{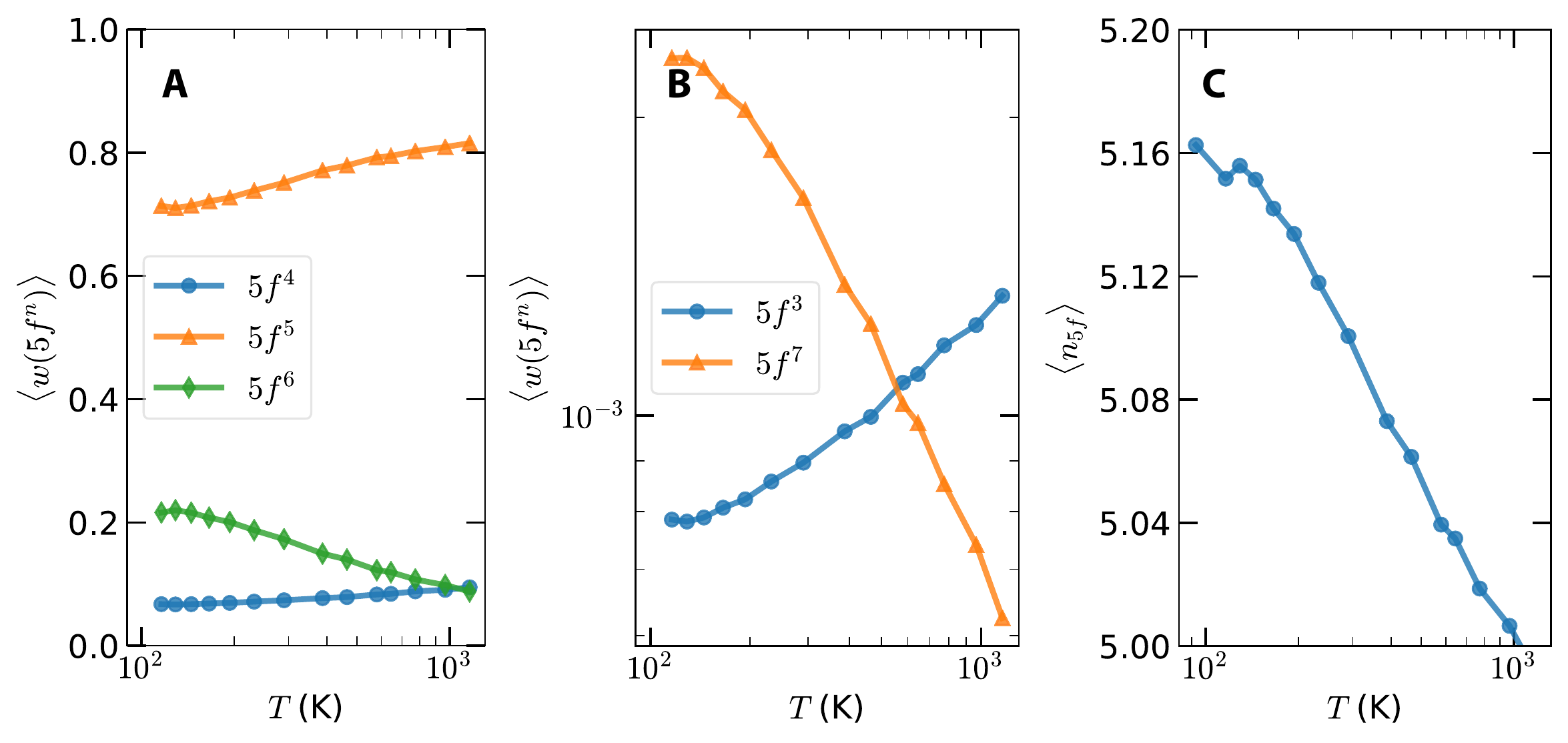}
\caption{\textbf{Temperature-dependent valence state fluctuations and 5$f$ occupancies of cubic phase Pu$_{3}$Ga obtained by the DFT + DMFT method.} \textbf{a} Probabilities of $5f^{4}$, $5f^{5}$, and $5f^{6}$ electronic configurations as a function of temperature $T$. \textbf{b} Probabilities of $5f^{3}$ and $5f^{7}$ electronic configurations as a function of temperature $T$. \textbf{c} $T$-dependent 5$f$ occupancies. \label{fig:prob}}
\end{figure*}

\textbf{Quasiparticle band structures.} The quasiparticle band structure (or equivalently, momentum-resolved spectral function) $A(\mathbf{k},\omega)$, which includes the one-particle vertex correction (i.e. the electron self-energy function), is an ideal theoretical tool to observe directly the status of the valence electrons of correlated electron materials. It can be validated by the angle-resolved photoemission spectroscopy~\cite{RevModPhys.92.011002}. Here, the calculated $A(\mathbf{k},\omega)$ along some selected high-symmetry directions in the irreducible Brillouin zone, are illustrated in Fig.~\ref{fig:akw}(c)-(k). Clearly, the quasiparticle band structures of cubic phase Pu$_{3}$Ga are strongly temperature-dependent. When the temperature is low ($T \leq 640$~K), the 5$f$ electrons are itinerant and form coherent hybridized bands. In the vicinity of the Fermi level (just $30 \sim 40$~meV below the Fermi energy), there are well-defined and nearly flat quasiparticle bands, which are likely connected with the itinerant 5$f$ electrons~\cite{PhysRevB.101.125123}. These $5f$ quasiparticle bands strongly hybridize with the $spd$ conduction bands. Along the $X-\Gamma$ and $\Gamma-M$ lines in the Brillouin zone, prominent $c-f$ hybridizations can be observed. In addition, approximately 0.5~eV and 0.9~eV below the Fermi level, there are stripe-like patterns in the spectra. It is speculated that these parallel features stem from the contributions of the spin-orbit splitting $5f_{5/2}$ and $5f_{7/2}$ states, respectively. When the temperature is high ($T > 640$~K), the 5$f$ electrons become more and more localized. Consequently, the spectra that close to the Fermi energy come to be more and more incoherent (see the band structures near the $M$ point). The quasiparticle bands, the $c-f$ hybridizations, and the stripe-like features disappear progressively. Thus, according to the calculated quasiparticle band structures, we can roughly conclude that there will be a temperature-driven itinerant-localized (or coherent-incoherent) crossover for the 5$f$ electrons in the cubic phase Pu$_{3}$Ga. The coherent temperature $T_{\text{coh}}$, which marks emergence of the quasiparticle bands and onset of the $c-f$ hybridizations, is around $640$~K. Finally, we would like to note that the positions of the quasiparticle bands and stripe-like features are hardly affected by the temperature in the low-temperature coherent states. 

\textbf{Fermi surface topology.} The evolution of 5$f$ correlated electronic states will manifest itself in the Fermi surface topology~\cite{PhysRevB.101.245156,PhysRevB.88.125106}. Fig.~\ref{fig:fs3d} shows the three-dimensional Fermi surfaces of cubic phase Pu$_{3}$Ga for three representative temperatures. There are three double degenerated bands crossing the Fermi level (No. of bands: 26 and 27, 28 and 29, and 30 and 31), which are labelled by using the Greek letters $\alpha$, $\beta$, and $\gamma$, respectively. The Fermi surfaces formed by the $\alpha$ bands are electron-like and look like distorted ellipsoids. When the temperature is increased, their topologies are not changed, but their volumes are increased. On the other hand, the $\beta$ and $\gamma$ bands forms extremely complicated Fermi surfaces with multiple sheets. Not only their volumes, but also their topologies will be affected by the temperature effect. To corroborate this statement, we further visualize the two-dimensional Fermi surfaces ($k_x-k_y$ planes) in Fig.~\ref{fig:fs2d}. The key lies in the $\beta$ bands. When $T \leq 190$~K, they intersect the $\Gamma-X$ line and form ellipsoid-like Fermi surfaces [see also Fig.~\ref{fig:fs3d}(c)], just like the $\alpha$ bands. However, when $T \ge 290$~K, they intersect the $M-X$ line instead and their Fermi surface topologies are completely changed. It means that between 190~K and 290~K, there must be an electronic Lifshitz transition for the 5$f$ correlated electronic states in the cubic phase Pu$_{3}$Ga, which can be detected by quantum oscillation experiments~\cite{PhysRevB.88.125106}.   

\textbf{Density of states.} Fig.~\ref{fig:dos}(a)-(c) show the total density of states $A(\omega)$ and $5f$ partial density of states $A_{5f}(\omega)$ of cubic phase Pu$_{3}$Ga at three representative temperatures. These spectra share some common characteristics. First of all, there are strong quasiparticle resonance peaks pinning at the Fermi level, which are associated with the nearly flat quasiparticle bands as already observed in the momentum-resolved spectral functions $A(\mathbf{k},\omega)$ (see Fig.~\ref{fig:akw} as well). Second, at -0.5~eV and -0.9~eV, two discernible satellite peaks exist. They are labelled as ``P1'' and ``P2'', respectively, in Fig.~\ref{fig:dos}(b) and (c) [these peaks at extremely high temperatures are not as obvious as those at low temperatures]. Once again, they are associated with the parallel stripe-like features seen in $A(\mathbf{k},\omega)$. Note that this three-peaks structure (including ``P1'', ``P2'', and the central quasiparticle peak) that lies in the occupied states is quite common in Pu~\cite{shim:2007,PhysRevB.101.125123,PhysRevB.99.125113} and its intermetallic compounds, such as PuTe~\cite{PhysRevB.81.035105}, PuCoIn$_{5}$~\cite{zhu:57001}, and PuCoGa$_{5}$~\cite{PhysRevB.98.035143,PhysRevB.87.020505}. In Ref.~[\onlinecite{PhysRevB.81.035105}], Chuck-Hou Yee \emph{et al.} named these peaks as quasiparticle multiplets. They proposed that the 5$f$ valence fluctuations in conjunction with the atomic multiplet structures, which may be remarkable in some plutonium-based intermetallic compounds, could drive the manifestation of a multiplet of many-body quasiparticle peaks. On the other hand, A. B. Shick \emph{et al.} have suggested another analogous picture. They tried to calculate the electronic structures of $\delta$-Pu and PuB$_{6}$ with the DFT + DMFT method~\cite{PhysRevB.87.020505,Shick2015}. They found that the obtained spectral functions can be crudely split into two distinctive parts. Near the Fermi level, the well-pronounced atomic multiplets structures dominate. However, in the high-energy regime, the multiplets are merged into broad lower and upper Hubbard bands. So, they called the materials that manifest similar features the ``Racah materials''. In other words, $\delta$-Pu belongs to the ``Racah metal'', while PuB$_{6}$ is the ``Racah semiconductor'' or ``Racah insulator''~\cite{PhysRevB.87.020505,Shick2015}. According to their definitions, we believe that the cubic phase Pu$_{3}$Ga belongs to the so-called Pu-based Racah metal. Third, there is a doublet of reflected peaks above the Fermi level. The positions of these peaks are approximately 0.4~eV and 0.8~eV. They are labelled as ``P3'' and ``P4'' in Fig.~\ref{fig:dos}(b)-(c). Chuck-Hou Yee \emph{et al.} have predicted similar peaks in plutonium chalcogenides through a slave-boson analysis~\cite{PhysRevB.81.035105}. So, as a whole, the fingerprint for the 5$f$ photoemission spectra of cubic phase Pu$_{3}$Ga includes five peaks, i.e. a central quasiparticle peak and four satellites. The positions of these satellites are actually symmetric about the central quasiparticle peak. 

Due to the spin-orbit coupling, the 5$f$ orbitals are split into six-fold degenerated $5f_{5/2}$ and eight-fold degenerated $5f_{7/2}$ states, respectively~\cite{RevModPhys.81.235,Brito:2020,PhysRevB.101.125123}. So, in order to determine the orbital characters of the five characteristic peaks, we further calculated the $j$-resolved 5$f$ partial density of states ($j = 5/2$ or $7/2$). The results are depicted in Fig.~\ref{fig:dos}(d)-(f). Clearly, the central quasiparticle peak is mainly composed of the $5f_{5/2}$ state. The $5f_{7/2}$ state remains insulating-like and manifests a gap in the Fermi level. The ``P1'' peak is largely from the $5f_{5/2}$ state. While for the other satellite peaks (``P2'', ``P3'', and ``P4''), they are built from a mixture of the $5f_{5/2}$ and $5f_{7/2}$ states.

Next, let us concentrate on the temperature dependence of these peaks. Fig.~\ref{fig:dos}(g) shows the temperature-dependent central quasiparticle peak. When the temperature is low, it should be a sharp and intense peak, indicating the itinerancy of the 5$f$ electrons. On the contrary, this peak should be suppressed and finally disappear with increasing temperature, which signals localization of the 5$f$ electrons and decay of the quasiparticles. Fig.~\ref{fig:dos}(h) shows spectral weights of the central quasiparticle peak, ``P1'', ``P2'', and $A_{5f}(\omega)$ at $\omega = 0$ as a function of temperature. All of them decrease monotonically with respect to temperature. Notice that the spectral weights of ``P3'' and ``P4'' show the same trends (not shown in this figure). The strong temperature dependence of the five peaks implies that all of them originate from quasiparticle resonances~\cite{PhysRevLett.101.056403,PhysRevB.81.035105}. In a short, we find that with the increasing temperature, an orbital selective 5$f$ itinerant-localized crossover would occur in the cubic phase Pu$_{3}$Ga. And at the same time, the quasiparticle multiplets collapse, which should be orbital selective as well. 

\textbf{Self-energy functions.} In principle, most of the electron correlation effects in correlated electron materials are encoded in their electron self-energy functions~\cite{RevModPhys.78.865,RevModPhys.68.13}. Thus, we can gain deep insight about electron correlations from them. In Fig.~\ref{fig:sig}(a)-(b), $Z|\text{Im}\Sigma(\omega)|$ for the $5f_{5/2}$ and $5f_{7/2}$ states are shown. Here, $Z$ denotes the quasiparticle weight or renormalization factor, which signals the strength of electron correlation and can be calculated from $\text{Re}\Sigma(\omega)$ through Eq.~(\ref{eq:renor})~\cite{RevModPhys.68.13}. So, $Z|\text{Im}\Sigma(\omega)|$ is the renormalized imaginary part of self-energy functions and $Z|\text{Im}\Sigma(0)|$ can be regarded as a reminiscence of the low-energy electron scattering rate~\cite{PhysRevB.99.125113}. When $T$ is small, both $Z|\text{Im}\Sigma_{5f_{5/2}}(0)|$ and $Z|\text{Im}\Sigma_{5f_{7/2}}(0)|$ approach zero. When $T$ is high, they become finite values and increase quickly with respect to temperature. At high-energy regime ($|\omega| > 1.0$~eV), the enhancement of $Z|\text{Im}\Sigma_{5f_{7/2}}(\omega)|$ is much stronger than that of $Z|\text{Im}\Sigma_{5f_{5/2}}(\omega)|$, which leads to a stronger suppression for the itinerancy of $5f_{7/2}$ state and explains its incoherent nature. Actually, the $5f_{7/2}$ state stays at the localized side even at 116~K [see Fig.~\ref{fig:dos}(f)], in contrast to the delocalized $5f_{5/2}$ state. This is another concrete evidence for the orbital differentiation of Pu-$5f$ electrons. 

From the self-energy functions, we can determine the coherent temperature quantitatively. In order to achieve this goal, we plot the imaginary parts of self-energy functions at the Fermi energy for the $5f_{5/2}$ and $5f_{7/2}$ states, i.e. $-\text{Im}\Sigma_{5f_{5/2}}(\omega = 0)$ and $-\text{Im}\Sigma_{5f_{7/2}}(\omega = 0)$, in Fig.~\ref{fig:sig}(c)-(d). Their first derivatives with respect to temperature, -$d\text{Im}\Sigma_{5f_{5/2}}(\omega = 0) / d\text{ln}T$ and -$d\text{Im}\Sigma_{5f_{7/2}}(\omega = 0) / d\text{ln}T$, are also plotted in the same figure. According to the literatures, the latter resembles the quasiparticle scattering rate. Its peak corresponds to the coherent temperature $T_{\text{coh}}$~\cite{Shim1615,PhysRevLett.123.217002}. When $T < T_{\text{coh}}$, the electron coherence sets in and heavy electron states appear, which lead to the rapid development of the quasiparticle peak near the Fermi level~\cite{RevModPhys.92.011002}. Anyway, $T_{\text{coh}}$ is about 700~K for the $5f_{5/2}$ state, which is coarsely consistent with the value that deduced from the spectral weights of the corresponding quasiparticle peaks [please refer to Fig.~\ref{fig:dos}(g) and (h)]. As for the $5f_{7/2}$ state, its coherent temperature is quite low ($T_{\text{coh}} \approx 100 $~K). It seems that the crossovers from the incoherent (localized) states to coherent (itinerant) states for the $5f_{5/2}$ and $5f_{7/2}$ states do not take place concurrently. There must be a wide range of temperature ($T \in [T_{\text{coh}}(5f_{7/2}), T_{\text{coh}}(5f_{5/2})]$), in which the 5$f$ electrons in the $5f_{5/2}$ state are coherent (itinerant) while those in the $5f_{7/2}$ state remain incoherent (localized).

We also utilized Eq.~(\ref{eq:renor}) to calculate the effective electron masses $m^{*}$ for the $5f_{5/2}$ and $5f_{7/2}$ states in cubic phase Pu$_{3}$Ga~\cite{RevModPhys.68.13}. The results are depicted in Fig.~\ref{fig:sig}(e). The temperature-dependent renormalized masses for the two states manifest quite different behaviors. With the increment of temperature, at first $m^{*}(5f_{5/2})$ increases steadily. Then it reaches its maximum value at approximately $T_{\text{coh}}(5f_{5/2})$. At last it decreases quickly. However, $m^{*}(5f_{7/2})$ decreases monotonically upon temperature. It finally approaches to a saturated value at high temperature.     

\textbf{Atomic eigenstate probabilities.} $\delta$-Pu is a typical mixed-valence metal, in which the ground state is a mixture of various $5f^{n}$ electronic configurations, so that the resulting 5$f$ occupancy deviates dramatically from the nominal value 5.0~\cite{shim:2007,Janoscheke:2015,PhysRevB.101.125123}. As is mentioned before, the cubic phase Pu$_{3}$Ga and $\delta$-Pu share some common features in their spectral functions, such as the famous three-peaks structure. So a question is naturally raised. Is the cubic phase Pu$_{3}$Ga a mixed-valence metal as well? In order to answer this question, we have to quantify the probability (or time) that 5$f$ electrons could stay (or spend) in each atomic eigenstate, which is computed by projecting the DMFT ground state onto the $5f$ electron atomic eigenstates~\cite{PhysRevB.81.035105}. Supposed that $p_{\Gamma}$ is the probability for the atomic eigenstate $|\Gamma \rangle$ and $n_{\Gamma}$ is the number of electrons, then the $5f$ valence can be deduced by $ \langle n_{5f} \rangle = \sum_{\Gamma} p_{\Gamma} n_{\Gamma} $, and the probability of $5f^{n}$ electronic configuration can be defined as $\langle w(5f^{n}) \rangle = \sum_{\Gamma} p_{\Gamma} \delta(n-n_{\Gamma})$.

In Fig.~\ref{fig:prob}, the $T$-dependent $\langle n_{5f} \rangle$ and $\langle w(5f^n) \rangle$ are shown. Please be aware that only the data for $5f^{3} \sim 5f^{7}$ configurations are plotted in this figure. The data for the other electronic configurations are too trivial to be shown. In low-temperature regime, it is no doubt that the cubic phase Pu$_{3}$Ga is a mixed-valence metal. Though the $5f^{5}$ configuration is overwhelming, the contributions from the $5f^{4}$ and $5f^{6}$ configurations are remarkable. Clearly, strong valence fluctuation should play a key role in generating the photoemission triplet below the Fermi level, and regulating the effective 5$f$ valence electrons. We obtained $\langle n_{5f} \rangle \approx 5.16$, which is close to those of $\delta$-Pu~\cite{shim:2007,PhysRevB.101.125123} and PuTe~\cite{PhysRevB.81.035105}. In contrast, this scenario is not valid any more in the high-temperature regime. At a first glance, the proportions of $5f^{5}$ and $5f^{4}$ configurations are increased slightly, while the one of $5f^{6}$ configuration is lowered. It implies that the valence state fluctuation is restrained and the 5$f$ valence electrons tend to spend more time in the $5f^{5}$ and $5f^{4}$ configurations. More important, the high temperature effect will shift the atomic multiplet energies (the energies of $5f^{6}$ states are raised relative to those of the $5f^{5}$ states), rendering the valence fluctuation too costly, thereby suppressing the quasiparticle multiplets. Finally, the 5$f$ electrons become more and more localized, and the cubic phase Pu$_{3}$Ga turns into an integral-valence metal ($\langle n_{5f} \rangle$ approaches its nominal value 5.0).

\begin{figure*}[t!]
\includegraphics[width=\textwidth]{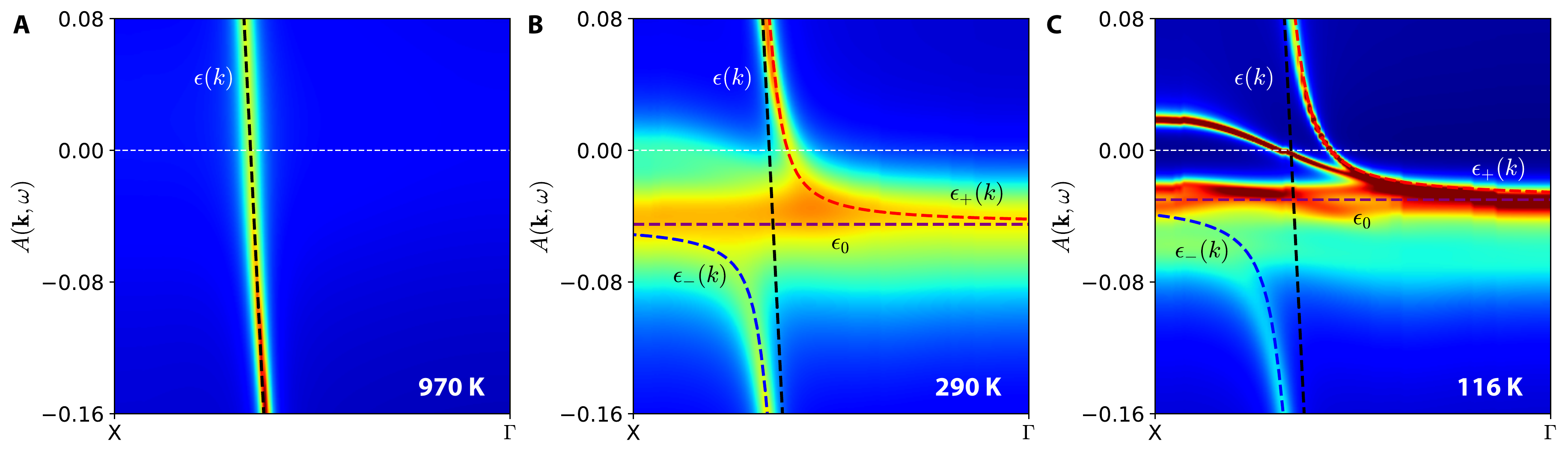}
\caption{\textbf{Temperature-dependent quasiparticle band structures (along the $X-\Gamma$ direction) of cubic phase Pu$_{3}$Ga obtained by the DFT + DMFT method.} \textbf{a} $T = 970$~K. \textbf{b} $T = 290$~K. \textbf{c} $T = 116$~K. The white horizontal dashed lines denote the Fermi level. Here, we utilized a periodical Anderson model [see Eq.~(\ref{eq:pam})] to fit the low-energy band structures. The colorful dashed lines are the fitting results. See main text for more details. \label{fig:hyb}}
\end{figure*}

\section{Discussion}

\textbf{Evolution of hybridization gap.} As is already illustrated in Fig.~\ref{fig:akw}(c)-(k), there is strong hybridization between the (localized) 5$f$ bands and the conduction bands when $T < T_{\text{coh}}$. The $c-f$ hybridization will open a hybridization gap for the conduction bands. Since this gap is close to the Fermi level, the physical properties of Pu$_{3}$Ga should be affected heavily by it. Thus, it is important to determine the size of the hybridization gap and elaborate its temperature dependence. Phenomenologically, the low-energy hybridized bands can be well described by a simple mean-field hybridization band picture (i.e. the periodical Anderson model)~\cite{RevModPhys.92.011002}. Within this picture, the energy dispersions read:
\begin{equation}
\label{eq:pam}
E_{\pm}(k) = \frac{[\epsilon_0 + \epsilon(k)] \pm \sqrt{[\epsilon_0 - \epsilon(k)]^2 + 4|V_k|^2}}{2},
\end{equation}  
where $\epsilon_0$ means the renormalized $5f$ energy level, $\epsilon(k)$ is the unrenormalized band dispersion for conduction electrons, and $V_k$ is the strength of hybridization. In the left side of this equation, the ``+'' and ``-'' symbols mean the upper and lower branches of hybridized bands, respectively. In Fig.~\ref{fig:hyb}(a) the band dispersion data at $T = 970$~K are shown. At such a high temperature the hybridization is negligible (i.e. $|V_k| = 0$ and $\epsilon_0 = 0$), so Eq.~(\ref{eq:pam}) is simplified to $E_{\pm}(k) = \epsilon(k)$. Thus, we used the data at $T = 970$~K to calibrate $\epsilon(k)$ (see the black dashed lines in Fig.~\ref{fig:hyb}). The band structures at $T = 290$~K and $116$~K are shown in Fig.~\ref{fig:hyb}(b) and (c), respectively. A fit to Eq.~(\ref{eq:pam}) gives $\epsilon_0 = -45$~meV and $|V_k| = 100$~meV for $T = 290$~K, and $\epsilon_0 = -30$~meV and $|V_k| = 120$~meV for $T = 116$~K. Thus, the direct hybridization gaps ($\Delta \approx 2|V_k|$) are 200~meV and 240~meV at 290~K and 116~K, respectively. These data suggest that the high temperature effect will push the 5$f$ energy level away from the Fermi level, restrain the $c-f$ hybridization, and reduce the hybridization gap $\Delta$. Note that besides the hybridized bands studied here, there also exist hybridizations between the quasiparticle multiplets and conduction bands. So, multiple hybridization gaps will be opened below the coherent temperature. The temperature-driven hybridization dynamics of this materials is rather complex. We would like to leave this problem to the future.     

\textbf{Electronic heat capacity.} The heat capacity of materials can be expressed as $C_v(T) = \gamma T + \beta T^3$, where the linear term $\gamma T$ comes from the contribution of electronic part and $\gamma$ is called the linear specific heat coefficient. Within the Fermi-liquid theory, $\gamma$ is given by the following equation:
\begin{equation}
\gamma = \pi k^2_{B} \sum_{\alpha} \frac{A_{\alpha}(0)}{Z_{\alpha}},
\end{equation}
where $\alpha$ is the orbital index, $A_{\alpha}(0)$ is the spectral weight at the Fermi level, and $Z_{\alpha}$ is the orbital-resolved renormalization factor~\cite{PhysRevLett.101.056403,RevModPhys.68.13}. Usually, large $\gamma$ [$\gamma > 100$ mJ / (mol~K$^2$)] is regarded as an useful indicator to distinguish the heavy-fermion metals from normal metals. We ignored the contributions from conduction bands, and assumed that the system is in the Fermi-liquid state when $T < T_{\text{coh}}$. The calculated $\gamma$ is about 112 mJ / (mol~K$^2$) at $T = 80$~K, which is almost double the value of $\delta$-Pu [$\gamma = 64 \pm 3$~mJ / (mol~K$^2$) as $T \to 0$~K]~\cite{PhysRevLett.91.205901}. Thus, we conclude that the cubic phase Pu$_{3}$Ga may be a promising candidate of Pu-based heavy-fermion system~\cite{bauer_review}.      

\textbf{Orbital differentiation and orbital selectivity.} As mentioned before, the 5$f$ orbitals will be split into $5f_{5/2}$ and $5f_{7/2}$ manifolds by the spin-orbit interaction~\cite{PhysRevB.101.125123,Brito:2020}. The band widths and effective interactions for the two kinds of bands are quite different. Such orbital differentiation is an ubiquitous feature for Pu and Pu-based materials~\cite{PhysRevB.101.125123,PhysRevB.99.125113,Brito:2020}. Not surprisingly, the calculated results for cubic phase Pu$_{3}$Ga  exhibit prominent orbital differentiation, such as in the $j$-resolved 5$f$ partial density of states, renormalization factors, and effective electron masses (see Fig.~\ref{fig:dos} and Fig.~\ref{fig:sig}). On the other hand, the orbital selectivity that manifested in the temperature-driven electronic structure transition is more important. Since the $5f_{5/2}$ manifold becomes incoherent at higher temperature than the $5f_{7/2}$ manifold, there are actually two coherent temperatures. In other words, when $T \in [T_{\text{coh}}(5f_{7/2}), T_{\text{coh}}(5f_{5/2})]$, the system falls in a specific situation, in which the $5f_{5/2}$ state retains itinerant while the $5f_{7/2}$ state becomes localized. We call it an orbital selective localized state~\cite{PhysRevB.99.045109}, which is analogous to the so-called orbital selective Mott phase as firstly discovered in $d$-electron systems. Additionally, the 5$f$ itinerant-localized crossovers in the cubic phase Pu$_{3}$Ga are orbital selective, which are reminiscent of the orbital selective Mott metal-insulator transitions~\cite{PhysRevLett.91.226401,PhysRevB.79.115119}.   

To summarize, we studied the electronic structures of cubic phase Pu$_{3}$Ga by using a state-of-the-art first-principles many-body approach. The temperature dependence of the correlated electronic states was detailed. When the temperature is increased, two itinerant-localized crossovers for Pu's 5$f$ electrons occur successively, giving rise to collapse of the quasiparticle multiplets, close of the hybridization gaps, and change in the Fermi surface topology. The calculated linear specific heat coefficient $\gamma$ is quite large [$\gamma > 100$ mJ / (mol~K$^2$)], which suggests a new candidate for the Pu-based heavy-fermion materials. Our calculated results not only provide a complete picture about how the 5$f$ correlated electronic states evolve with respect to temperature in Pu-based materials for the first time, but also shed new light onto the complex electronic structures of Pu-Ga system. Further studies about the other Pu-Ga intermetallic compounds will be undertaken.      

\section{Methods}

\textbf{DFT calculations.} The DFT calculations were performed by using the \texttt{WIEN2K} code, which implements a full-potential linearized augmented plane-wave formalism~\cite{wien2k}. We adopted the experimental crystal structure of Pu$_{3}$Ga ($a_0 = 4.0$~\AA) and ignored the thermal expansion~\cite{PhysRevB.96.134102}. The radius of muffin-tin spheres for Pu and Ga atoms were 2.5 au and 2.1 au, respectively. $R_{\text{MT}}K_{\text{MAX}} = 8.0$. The $5f$, $6d$, and $7s$ orbitals in Pu and $4f$, $5d$, and $6s$ orbitals in Ga were treated as valence states. The rests are treated as core states. The $k$-mesh for Brillouin zone integration was $15 \times 15 \times 15$. The generalized gradient approximation, namely the Perdew-Burke-Ernzerhof functional~\cite{PhysRevLett.77.3865}, was used to evaluate the exchange-correlation potential. The spin-orbit coupling effect was included in a variation manner. The system was assumed to be nonmagnetic~\cite{PhysRevB.96.134102,PhysRevB.100.184101}.

\textbf{DFT + DMFT calculations.} We utilized the \texttt{eDMFT} software package, which was developed by K. Haule \emph{et al.}, to perform the DFT + DMFT calculations~\cite{PhysRevB.81.195107}. The many-body nature of the Pu-$5f$ orbitals was captured by the DMFT formalism. The Coulomb interaction matrix for Pu-$5f$ orbitals was constructed by using the Slater integrals. The Coulomb repulsion interaction parameter $U$ and Hund's exchange interaction parameter $J_{\text{H}}$ are 6.0~eV and 0.7~eV, respectively, which were taken from Reference~[\onlinecite{PhysRevB.96.134102}] directly. We used the $|J, J_z\rangle$ basis to construct the local impurity Hamiltonian. A large energy window, from -10~eV to 10~eV with respect to the Fermi level, was used to build the DMFT projector, which was used to project the Kohn-Sham basis to local basis. The vertex-corrected one-crossing approximation (OCA) impurity solver~\cite{PhysRevB.64.155111} was employed to solve the resulting multi-orbital Anderson impurity models. In order to reduce the computational consumes and accelerate the calculations, we have to truncate the Hilbert space of the local impurity problems. Only contributions from those atomic eigenstates with $N \in [3,7]$ were retained during the calculations. The calculated results were also crosschecked by using the numerically exact and unbiased hybridization expansion version continuous-time quantum Monte Carlo impurity solver (dubbed CT-HYB)~\cite{PhysRevB.75.155113,RevModPhys.83.349}. The double-counting term is used to cancel out the excess amount of electronic correlations that is already partly included in the DFT part. In the present work, we just selected the fully localized limit scheme to describe the double-counting term~\cite{jpcm:1997}. It reads 
\begin{equation}
\Sigma_{\text{dc}} = U \left(n_{5f} - \frac{1}{2}\right) - \frac{ J_{\text{H}} } { 2} \left( n_{5f} -1 \right)
\end{equation}
where $n_{5f}$ is the nominal occupancy of Pu-$5f$ orbitals. It was fixed to be 5.0 during the DFT + DMFT calculations. We conducted charge fully self-consistent DFT + DMFT calculations. About $60 \sim 80$ DFT + DMFT iterations were enough to obtain good convergence. The convergent criteria for charge density and total energy were $10^{-4}$~e and $10^{-5}$~Ry, respectively.

\textbf{Analytical continuation.} Since the OCA impurity solver works at real axis directly~\cite{PhysRevB.64.155111}. It doesn't need to do tedious analytical continuation for the self-energy functions. However, once the CT-HYB impurity solver is used, the output self-energy functions are usually defined at Matsubara frequency axis. They cannot be used to calculate physical observables directly. The obtained Matsubara self-energy functions $\Sigma(i\omega_n)$ were analytically continued by using the maximum entropy method at first~\cite{jarrell}. Then the resulting self-energy functions at real axis $\Sigma(\omega)$ were applied to calculate the momentum-resolved spectral functions $A(\mathbf{k},\omega)$ and density of states $A(\omega)$ via the following equations:
\begin{equation}
A(\omega) = \int_{\Omega}  A(\mathbf{k},\omega) \text{d} \mathbf{k},
\end{equation}
and
\begin{equation}
\label{eq:akw}
A(\mathbf{k},\omega) = -\frac{1}{\pi}\text{Im}\frac{1}{(\omega + \mu) \hat{\mathbf{I}} - \hat{H}_{\text{KS}}(\mathbf{k}) - \hat{E}(\mathbf{k})[ \Sigma(\omega) - \Sigma_{\text{dc}}]}.
\end{equation}
Here $\hat{\mathbf{I}}$ is the identity matrix, $\mu$ the chemical potential, $\hat{H}_{\text{KS}}(\mathbf{k})$ the Kohn-Sham Hamiltonian, $\hat{E}(\mathbf{k})$ the embedding projector in momentum space~\cite{PhysRevB.81.195107}. Finally, $\Sigma(\omega)$ were used to calculate quasiparticle weight $Z$ and effective electron masses $m^{*}$ for Pu-$5f$ electrons~\cite{RevModPhys.68.13}:
\begin{equation}
\label{eq:renor}
Z^{-1} = \frac{m^{*}}{m_e} = 1 - \frac{\partial \text{Re}\Sigma(\omega)}{\partial \omega}\Big|_{\omega = 0},
\end{equation}
where $m_e$ means the mass of bare (non-interaction) electron.

\section{Data availability}

The data that support the findings of this study will be made available upon reasonable requests to the corresponding author (L.H.).

\section{Code availability}

The computer codes used in this study can be obtained from the following websites:
\begin{itemize}
\item \texttt{WIEN2K}~http://susi.theochem.tuwien.ac.at
\item \texttt{eDMFT}~http://hauleweb.rutgers.edu/tutorials/
\end{itemize}

\section{Acknowledgements}

This work is supported by the National Natural Science Foundation of China (under Grants No.~11874329, No.~11934020, and No.~11704347), and the Science Challenge Project of China (under Grant No.~TZ2016004).

\section{Author contributions}

L. H. supervised the project, carried out the DFT + DMFT (OCA) calculations, and analyzed the results. H.Y.L. carried out the DFT + DMFT (CT-HYB) calculations. L.H. wrote the paper with contributions and comments from H.Y.L.

\section{Competing interests}

The authors declare no competing interests.

\bibliography{Pu3Ga}

\end{document}